\documentclass[article,shortnames]{jss}
\usepackage{amsmath}
\usepackage{amssymb}
\usepackage{bm}

\author{Peter Carbonetto\\University of Chicago \And
  Xiang Zhou\\University of Michigan \And
  Matthew Stephens\\University of Chicago}

\title{\pkg{varbvs}: Fast Variable Selection for Large-scale Regression}

\Plainauthor{Peter Carbonetto, Xiang Zhou, Matthew Stephens}

\Plaintitle{varbvs: Fast Variable Selection for Large-scale Regression}

\Shorttitle{\pkg{varbvs}: Fast Variable Selection for Large-Scale
  Regression}

\Abstract{We introduce \pkg{varbvs}, a suite of functions written in
  \proglang{R} and \proglang{MATLAB} for regression analysis of 
  large-scale data sets using Bayesian variable selection methods.
  We have developed numerical optimization algorithms based on
  variational approximation methods that make it feasible to apply
  Bayesian variable selection to very large data sets.
  With a focus on examples from genome-wide association studies, we
  demonstrate that \pkg{varbvs} scales well to data sets with hundreds
  of thousands of variables and thousands of samples, and has features
  that facilitate rapid data analyses. Moreover, \pkg{varbvs} allows
  for extensive model customization, which can be used to incorporate
  external information into the analysis.
  We expect that the combination of an easy-to-use interface and
  robust, scalable algorithms for posterior computation will encourage
  broader use of Bayesian variable selection in areas of applied
  statistics and computational biology. The most recent \proglang{R}
  and \proglang{MATLAB} source code is available for download at
  Github (\url{https://github.com/pcarbo/varbvs}),
  and the \proglang{R} package can be installed from CRAN
  (\url{https://cran.r-project.org/package=varbvs}).}


\Plainkeywords{Bayesian variable selection, linear regression,
  logistic regression, approximate posterior computation, variational
  inference, Bayes factors, genome-wide association studies,
  quantitative trait locus mapping, R, MATLAB}

\Address{
  Peter Carbonetto\\
  Research Computing Center\\
  and Department of Human Genetics\\
  University of Chicago\\
  Chicago, Illinois, USA 60637\\
  E-mail: \email{pcarbo@uchicago.edu}\\
  URL: \url{http://github.com/pcarbo}
}

\begin{document}

\section{Introduction}

Bayesian variable selection (BVS) models, and extensions to these
models, have recently been shown to provide attractive solutions to a
number of important problems in genome-wide association studies (e.g.,
\citealt{carbonetto-2012, carbonetto-2013, guan-2011, lee-2008,
  hoggart-2008, logsdon-2010, meuwissen-2001, moser-2015,
  wallace-2015, zhou-2013}).
%
%
%
Despite this progress, BVS methods have not been widely adopted for
genome-wide association studies (GWAS) and other areas where
large-scale regression is applied. One limiting factor is that
computing exact posterior probabilities, which reduces to a
high-dimensional integration problem, is intractable except in very
small data sets, and standard approaches for approximating these
high-dimensional integrals using Monte Carlo techniques scale poorly
to large data sets \citep{bottolo-2010, clyde-2011, dellaportas-2002,
  erbe-2012, guan-2011, perez-2014, wallace-2015, zhou-2013}.
%
%
%
A second barrier is that the choice of priors requires considerable
expertise in Bayesian data analysis. We aim to address these
limitations and make BVS methods more accessible.

Here, we present a software toolkit for fitting variable selection
models to large-scale data sets. We call our software
\pkg{varbvs}---short for ``variational Bayesian variable
selection''---as it builds on Bayesian models for variable selection
in regression \citep{george-1993, mitchell-1988, ohara-2009} and
variational approximation techniques for fast posterior computation
\citep{blei-2016, jordan-1999, logsdon-2010, ormerod-2010,
  wainwright-2008}. We have developed efficient implementations for
both \proglang{R} \citep{R} and \proglang{MATLAB} \citep{matlab},
which we have applied to data sets containing hundreds of thousands of
variables and thousands of samples. \pkg{varbvs} also provides default
priors that are suitable for many problem areas, while allowing
for extensive customization. While our initial motivation was to
facilitate use of multiple regression models for genome-wide
association studies \citep{carbonetto-2012, guan-2011}, Bayesian
variable selection methods are general and widely applicable, and we
expect that \pkg{varbvs} will be useful in many other areas of applied
statistics and computational biology.

Our second aim is to provide an alternative to commonly used toolkits
for penalized regression.
%
%
\pkg{varbvs} is comparable to the popular \proglang{R} package
\pkg{glmnet} \citep{glmnet}, which combines penalized sparse
regression---specifically, the Lasso \citep{tibshirani-1994} and the
Elastic Net \citep{zou-2005}---with advanced optimization techniques
\citep{friedman-2007}.
The \pkg{varbvs} interface is designed to be similar to \pkg{glmnet}
so that researchers already familiar with these methods can easily
explore the benefits of the BVS approach. In our first example
(Sec.~\ref{sec:leukemia}), we illustrate the shared features and
differences of \pkg{glmnet} and \pkg{varbvs}.

%
An important advantage of BVS over penalized regression is that it
provides a measure of uncertainty in the parameter estimates. For
example, \pkg{varbvs} computes, for each candidate variable, the
probability that the variable is included in the regression
model---what we call the ``posterior inclusion probability'' (PIP).
A second advantage of BVS over penalized regression is that it allows
for the possibility of model comparison through approximate
computation of Bayes factors \citep{kass-1995}. We demonstrate both
advantages in the examples below.

The structure of the paper is as follows. In Sec.~\ref{sec:leukemia},
we given an extended example that illustrates the key features of
\pkg{varbvs}, comparing it to \pkg{glmnet}. Section~\ref{sec:bvs}
briefly reviews Bayesian variable selection in regression, and
explains how it is implemented in \pkg{varbvs}.
Sections \ref{sec:cfw} and \ref{sec:cd} give more advanced examples
illustrating the application of \pkg{varbvs} to large data sets with
tens or hundreds of thousands of variables.
In Section~\ref{sec:discussion}, we end with additional discussion and
recommendations on applying \pkg{varbvs} to small and large data sets.

Although this paper focuses on the \proglang{R} package,
we note that a \proglang{MATLAB} interface is also available. The
\proglang{MATLAB} implementation can be substantially faster for large
data sets thanks to \proglang{MATLAB}'s state-of-the-art numerical
computing platform. For this reason, we use the \proglang{MATLAB}
interface for the large data analyses in Sections \ref{sec:cfw} and
\ref{sec:cd}.

\section[Example illustrating features of glmnet and varbvs]{Example
  illustrating features of \pkg{glmnet} and \pkg{varbvs}}
\label{sec:leukemia}

We illustrate \pkg{glmnet} and \pkg{varbvs} on a smaller data set that
has been used in previous papers to compare methods for penalized
regression (e.g., \citealt{breheny-2011, glmnet, tibshirani-2005,
  zou-2005}). Our example is meant to demonstrate the \pkg{varbvs}
\proglang{R} interface, and to provide some intuition for the different
properties of BVS and penalized regression as implemented by
\pkg{varbvs} and \pkg{glmnet}, respectively. 
The ``leukemia'' vignette in the \proglang{R} package reproduces the
results and figures in this section.

The data consist of expression levels recorded for 3,571 genes in 72
patients with leukemia \citep{golub-1999}. The genes are the candidate
variables. The binary outcome, modeled using a logistic regression,
encodes the disease subtype: acute lymphobastic leukemia (ALL) or
acute myeloid leukemia (AML). We use the preprocessed data of
\cite{dettling-2004} retrieved from the supplementary materials
accompanying \cite{glmnet}. The data are represented as a $72 \times
3571$ matrix \code{X} of gene expression levels, and a vector \code{y}
of 72 binary disease outcomes. We fit logistic models to these data
using \pkg{glmnet} and \pkg{varbvs}, and explore properties of the
fitted models.

We begin with \pkg{glmnet}. For each setting of the penalty strength
parameter $\lambda$, \pkg{glmnet} fits a logistic regression by
solving this convex optimization problem:
\begin{equation}
\underset{\beta_0 \,\in\, \mathbb{R}, \beta \,\in\, \mathbb{R}^p}
         {\mathrm{minimize}} \quad
- \frac{1}{n} \sum_{i=1}^n \Pr(y_i \,|\, x_i, \beta_0, \beta)
+ \frac{\lambda}{2} (1 - \alpha) \|\beta\|_2^2
+ \lambda \alpha \|\beta\|_1,
\label{eq:glmnet}
\end{equation}
where $x_i$ is the vector of expression levels recorded in patient
$i$, $y_i$ is the disease outcome, $n = 72$ is the number of samples,
$p = 3571$ is the number of candidate variables, $\beta$ is the vector
of logistic regression coefficients, $\beta_0$ is the intercept,
$\|\,\cdot\,\|_1$ is the $\ell_1$-norm, $\|\,\cdot\,\|_2$ is the
Euclidean ($\ell_2$) norm, $\Pr(y_i \,|\, x_i, \beta_0, \beta)$ is the
logistic regression likelihood (see
Equation~\ref{eq:logistic-regression} below). Following \cite{glmnet},
$\lambda$ determines the overall penalty strength, and $\alpha$
balances the $\ell_1$ and $\ell_2$ penalty terms (here, we set $\alpha
= 0.95$). 

This model fitting is accomplished with a single call to the
\code{glmnet} function:

\begin{Schunk}
\begin{Sinput}
R> data(leukemia, package = "varbvs")
R> library(glmnet)
R> X <- leukemia$x
R> y <- leukemia$y
R> colnames(X) <- paste0("X", 1:3571)
R> fit.glmnet <- glmnet(X, y, family = "binomial", alpha = 0.95, 
+      lambda = 10^(seq(0, -2, -0.05)))
\end{Sinput}
\end{Schunk}

(Note that we overrode the default \code{lambda} to make the plots
below easier to follow---it yields a similar result to the default
setting.) As part of the \pkg{glmnet} model fitting, the intercept and
regression coefficients are estimated for each entry of \code{lambda},
and these are represented as a $3572 \times 42$ matrix
\code{coef(fit.glmnet)}.

\begin{figure}[t!]
\begin{center}
\includegraphics{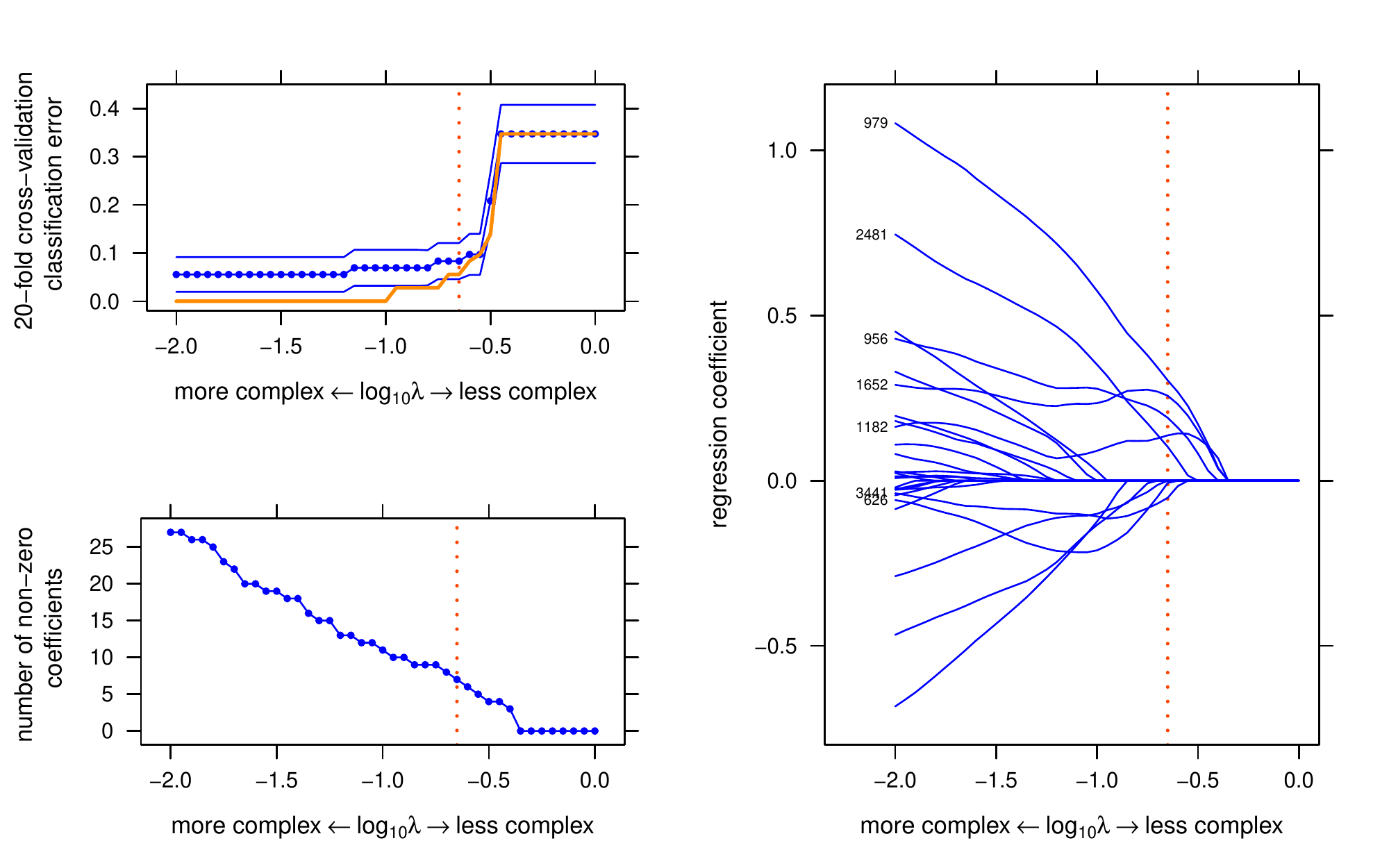}
\caption{\pkg{glmnet} analysis of leukemia data. {\em top-left panel:}
  $\ell_1$-penalty strength parameter ($\lambda$) against proportion
  of left-out samples in 20-fold cross-validation that are
  misclassified by the Elastic Net model. Top and bottom curves
  give confidence intervals for the classification error across the 20
  folds; middle curve in blue is the mean classification error. The
  wider orange line gives the classification error for the model
  fitted to the entire data set using function \code{glmnet}, which is
  added for comparison to the \pkg{varbvs} results. {\em bottom-left
    panel:} Number of variables included in model (variables with
  non-zero coefficients) at each $\lambda$ setting. {\em right-hand
    panel:} Regression coefficients at each setting of $\lambda$. The
  labeled curves are the 7 variables included in the model at $\lambda
  = 0.224$, the setting chosen by cross-validation (dotted vertical
  red lines).}
\label{fig:glmnet-leukemia} 
\end{center}
\end{figure}

The right-hand plot in Fig.~\ref{fig:glmnet-leukemia} shows the
characteristic shrinkage pattern of sparse regression methods such as
the Lasso and the Elastic Net; as $\lambda$ becomes larger, the
$\ell_1$-penalty term becomes more prominent, thereby encouraging more
shrinkage of the regression coefficients. The bottom-left plot shows
the total number of variables
with non-zero coefficients at each $\lambda$, and is another way
visualizing this shrinkage pattern.

The top-left plot in Fig.~\ref{fig:glmnet-leukemia} shows the
evolution of the cross-validation classification error at the same
settings of $\lambda$. Small values of $\lambda$ allow for more
complex models, and therefore offer a better fit to the data.
To guard against overly complex models that ``overfit'' to the data,
\pkg{glmnet} uses cross-validation:

\begin{Schunk}
\begin{Sinput}
R> out.cv.glmnet <- cv.glmnet(X, y, family = "binomial", type.measure = "class", 
+      lambda = 10^(seq(-2, 0, 0.05)), alpha = 0.95, nfolds = 20)
R> print(out.cv.glmnet$lambda.1se)
\end{Sinput}
\begin{Soutput}
[1] 0.2239
\end{Soutput}
\end{Schunk}

\setkeys{Gin}{width=1.025\textwidth}

The penalty strength selected by 20-fold cross-validation,
\code{lambda.1se}, is depicted in the figure by the dotted vertical
red lines.
At this penalization level, \pkg{glmnet} yields a very sparse
regression model---only 7 out of the 3,571 gene expression features
are included in the model (Fig.~\ref{fig:glmnet-leukemia}, right-hand
panel)---yet these 7 features are sufficient to correctly predict the
leukemia outcome in 68 of the 72 training examples:

\begin{Schunk}
\begin{Sinput}
R> y.glmnet <- c(predict(fit.glmnet, X, s = out.cv.glmnet$lambda.1se, 
+      type = "class"))
R> print(table(true = factor(y), pred = factor(y.glmnet)))
\end{Sinput}
\begin{Soutput}
    pred
true  0  1
   0 47  0
   1  4 21
\end{Soutput}
\end{Schunk}

The entire \pkg{glmnet} analysis, including cross-validation, is very
fast; it took less than 3 seconds to run on a computer with a 1.86 GHz
Intel Core 2 Duo processor.

Next, we compare this \pkg{glmnet} analysis against an analysis of the
same data using \pkg{varbvs}. As before, we use logistic regression to
model the outcome given the regression coefficients. However, rather
than optimize the coefficients subject to a penalty, we introduce an
exchangeable ``spike-and-slab'' prior \citep{mitchell-1988,
  george-1993} on the coefficients $\beta$,
\begin{equation}
\Pr(\beta_i \,|\, \pi, \sigma_a^2) =
(1 - \pi) \delta_0 + \pi N(0, \sigma_a^2),
\label{eq:spike-and-slab}
\end{equation}
and we compute approximate posterior probabilities with respect to
this prior. Additionally, instead of a two-step analysis---modeling
fitting and cross-validation---the \pkg{varbvs} analysis is
accomplished in a single function call:

\begin{Schunk}
\begin{Sinput}
R> library(varbvs)
R> fit.varbvs <- varbvs(X = X, y = y, Z = NULL, family = "binomial")
\end{Sinput}
\end{Schunk}

This command took about 30 seconds to run on the same computer.

\begin{figure}[t!]
\begin{center}
\includegraphics{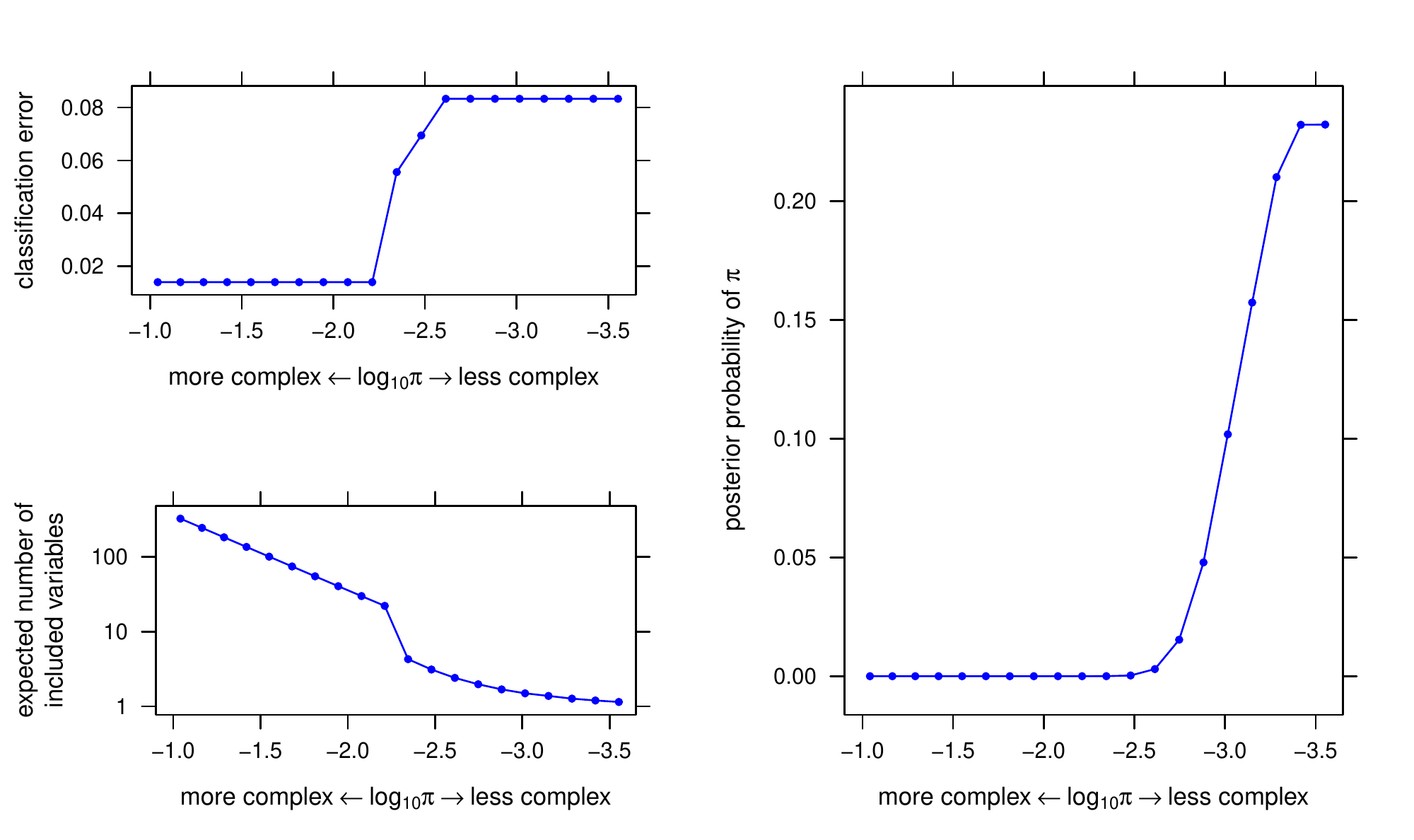}
\caption{\pkg{varbvs} analysis of leukemia data. {\em top-left panel:}
  Prior inclusion probability $\pi$ against proportion of samples that
  are misclassified by the \pkg{varbvs} model. This should be compared
  against the wider orange line in the top-left panel of
  Fig.~\ref{fig:glmnet-leukemia}. {\em bottom-left panel:} For each
  $\pi$ setting, expected number of variables (variables with non-zero
  coefficients) that are included in the model. {\em top-right panel:}
  Estimated posterior distribution of $\pi$.}
%
\label{fig:varbvs-leukemia} 
\end{center}
\end{figure}

The complexity of the regression model is controlled by the prior,
which is determined by two parameters: the prior probability $\pi$
that a variable is included in the regression model, and $\sigma_a^2$,
the prior variance of the non-zero regression coefficients.
Similar to \pkg{glmnet}, \pkg{varbvs} fits a model separately for each
setting of $\pi$. The default is a grid with 20 settings of $\pi$.
Parameter $\sigma_a^2$ is estimated separately for each setting of
$\pi$. This is only the default behaviour---it is also possible to
define a grid over $\pi$ and $\sigma_a^2$ and fit models across all
grid points.

To illustrate the effect that $\pi$ has on model complexity, we
compute the classfication error at each setting of $\pi$, stored as
the prior log-odds, $\log_{10}(\frac{\pi}{1-\pi})$, in
\code{fit.varbvs$logodds}:

\begin{Schunk}
\begin{Sinput}
R> m <- length(fit.varbvs$logodds)
R> err <- rep(0, m)
R> for (i in 1:m) {
+      r <- subset(fit.varbvs, logodds == fit.varbvs$logodds[i])
+      ypred <- predict(r, X)
+      err[i] <- mean(y != ypred)
+  }
\end{Sinput}
\end{Schunk}

The classification error is shown in the top-left panel of
Fig.~\ref{fig:varbvs-leukemia}. Like \pkg{glmnet}, the \pkg{varbvs}
model predictions improve as the variable selection prior allows for
more complex models. However, in contrast to \pkg{glmnet}, 
cross-validation is not needed to select an appropriate level of
regularization---the Bayesian inference approach automatically weighs
the accuracy of the model predictions against the model complexity
\citep{jeffreys-1992, mackay-1992}. In this example, more complex
models---{\em i.e.}, more variables included in the model---offer only
a marginally better fit to the data, so the posterior distribution is
most concentrated on less complex models
(Fig.~\ref{fig:varbvs-leukemia}, top-right). In fact, the posterior is
most concentrated on models in which the variance in the leukemia
outcome is largely explained by a single feature
(Fig.~\ref{fig:varbvs-leukemia}, bottom-left).

Like \pkg{glmnet}, the \pkg{varbvs} model also predicts the regression
outcomes with good accuracy:

\begin{Schunk}
\begin{Sinput}
R> y.varbvs <- predict(fit.varbvs, X)
R> print(table(true = factor(y), pred = factor(y.varbvs)))
\end{Sinput}
\begin{Soutput}
    pred
true  0  1
   0 44  3
   1  3 22
\end{Soutput}
\end{Schunk}

The accuracy of \pkg{varbvs} is statistically indistinguishable from
\pkg{glmnet} in this case (6 errors compared to 4 errors in the the
\pkg{glmnet} analysis) and this accuracy is achieved by concentrating
the posterior distribution on much simpler models than \pkg{glmnet} in
which the variance in the leukemia outcome is mostly explained by a
single predictor. This is possible in \pkg{varbvs} because the
shrinkage behaviour is quite different from \pkg{glmnet}; as $\pi$ is
decreased, the model becomes sparser (fewer included variables), but
the most strongly included variable is hardly shrunk at all. That is,
\pkg{varbvs} can achieve strong shrinkage of effects near zero without
correspondingly strong shrinkage of the important predictors. This is
a highly desirable feature that convex penalization methods such as
the Lasso and Elastic Net struggle to achieve.

By default, \pkg{varbvs} yields {\em averaged} predictions---that is,
the model predictions are collected from all hyperparameter settings,
and the final prediction \code{y.varbvs} is computed as a weighted
average of the individual predictions, with weights given by posterior
probabilities of the hyperparameter settings
(Fig.~\ref{fig:varbvs-leukemia}, right-hand side). 
Model averaging in BVS is typically computationally prohibitive in
large-scale data sets, but the variational approximation yields a
simple and efficient approach to account for uncertainty.

%
%

Finally, we note that parameter estimation in \pkg{varbvs} is a
nonconvex optimization problem (as we explain below), so it can be
sensitive to variable ordering and initialization of the fitting
procedure. By contrast, \pkg{glmnet} will always produce the same
model fit for the same data because the parameter estimation reduces
to a convex optimization problem. For example, if we reorder the
columns of \code{X} before fitting the \pkg{varbvs} model,

\begin{Schunk}
\begin{Sinput}
R> fit.varbvs.alt <- varbvs(X = X[, sample(3571)], y = y, Z = NULL, 
+      family = "binomial")
\end{Sinput}
\end{Schunk}

then the variance in the leukemia outcome is again is mostly explained
by two different variables:

\begin{Schunk}
\begin{Sinput}
R> print(summary(fit.varbvs, nv = 3)$top.vars)
\end{Sinput}
\begin{Soutput}
      index variable     prob PVE    coef  Pr(coef.>0.95)
X3441  3441    X3441 0.999949  NA -4.2957 [-5.072,-3.527]
X1608  1608    X1608 0.001876  NA -0.8076 [-1.430,-0.214]
X2529  2529    X2529 0.001291  NA  0.7560 [+0.165,+1.350]
\end{Soutput}
\begin{Sinput}
R> print(summary(fit.varbvs.alt, nv = 3)$top.vars)
\end{Sinput}
\begin{Soutput}
      index variable     prob PVE   coef  Pr(coef.>0.95)
X1182  1336    X1182 0.989419  NA 2.9051 [+2.145,+3.651]
X2507    46    X2507 0.989007  NA 2.3199 [+1.289,+3.011]
X2888  1104    X2888 0.001825  NA 0.6126 [+0.055,+1.124]
\end{Soutput}
\end{Schunk}

As expected, the top two variables in the second \pkg{varbvs} analysis
are strongly correlated with the top variable from the first analysis:

\begin{Schunk}
\begin{Sinput}
R> print(cor(X[, "X3441"], X[, c("X1182", "X2507")]))
\end{Sinput}
\begin{Soutput}
       X1182   X2507
[1,] -0.7717 -0.6737
\end{Soutput}
\end{Schunk}

More generally, when multiple variables are strongly correlated with
each other, the parameter estimation can be sensitive to the variable
ordering and initialization of optimization procedure. To ensure that
a \pkg{varbvs} analysis is reproducible, we recommend using
\code{set.seed} to fix the sequence of pseudorandom numbers, and
checking that different seeds and/or variable orderings yield
reasonably consistent estimates (see ``Summary and discussion'').

\section[Bayesian variable selection, and the varbvs R
  interface]{Bayesian variable selection, and the \pkg{varbvs}
  \proglang{R} interface}
\label{sec:bvs}

In this section, we define the general analysis setup: the regression
model (Sec.~\ref{sec:regression}), the variable selection priors
(Sec.~\ref{sec:prior}), and the approach taken to efficiently compute
posterior quantities (Sections \ref{sec:posterior} and
Sec.~\ref{sec:posterior-hyperparameters}). As we walk through the
setup, we connect aspects of the analysis to the \code{varbvs}
interface, then we review the interface in
Sec.~\ref{sec:interface}. For background on Bayesian approaches to
variable selection, see \cite{george-2000} and \cite{ohara-2009}.


\subsection{Regression model}
\label{sec:regression}

The data consist of an $n \times p$ matrix ${\bf X}$ containing
observations $x_{ij}$ of the candidate variables, an $n \times m$
matrix ${\bf Z}$ containing measurements $z_{ij}$ of the covariates,
and a vector $y = (y_1, \ldots, y_n)^T$ containing observations of the
regression outcome. These data are provided to function \code{varbvs}
through arguments \code{X}, \code{Z} and \code{y}.

The \pkg{varbvs} package implements methods for both linear regression
(\code{family = "gaussian"}) and logistic regression (\code{family =
  "binomial"}). For linear regression, the outcome $Y$ is modeled as a
linear combination of the candidate predictors, covariates and
residuals $\epsilon \sim N(0, \sigma^2)$:
\begin{equation}
Y = \sum_{i=1}^m Z_i u_i + \sum_{i=1}^p X_i \beta_i + \epsilon.
\label{eq:linear-regression}
\end{equation}
For logistic regression, we model the log-odds of $Y = 1$ as a linear
combination of the predictors and covariates:
\begin{equation}
\log\bigg\{\frac{\Pr(Y = 1)}{\Pr(Y = 0)}\bigg\} = 
\sum_{i=1}^m Z_i u_i + \sum_{i=1}^p X_i \beta_i.
\label{eq:logistic-regression}
\end{equation}
(Since $\sigma^2$ is not needed for logistic regression, in the
definitions below we set $\sigma^2 = 1$ in this case.)  At least one
covariate, the intercept, must always be included in the model.

\subsection{Variable selection prior}
\label{sec:prior}

We use the one of the most successful Bayesian approaches to variable
selection, based on the ``spike-and-slab'' prior (Equation
\ref{eq:spike-and-slab}).
%
Small values of $\pi$ encourage sparse regression models, in which
only a small proportion of the candidate variables $X_i$ help predict
the outcome $Y$.
The rationale for this prior has been given in previous papers (e.g.,
\cite{carbonetto-2012, guan-2011, servin-2007, zhou-2013}), and we do
not repeat this discussion here.


The grid of hyperparameter settings $(\sigma^2, \sigma_a^2, \pi)$ is
defined by three inputs to \code{varbvs}: \code{sigma}, the residual
variance for linear regression (for logistic regression, we set
$\sigma^2 = 1$); \code{sa}, the prior variance of the regression
coefficients; and \code{logodds}, the prior inclusion probability
$\pi$ defined on the log-odds scale, $\log_{10}
\big\{\frac{\pi}{1-\pi}\}$. A plausible range of prior log-odds are
generated automatically if they are not supplied by input
\code{logodds}. When inputs \code{sigma} and \code{sa} are not
provided, the default behaviour is to estimate these parameters
separately for each setting of $\pi$.
%

This standard variable selection prior (Equation \ref{eq:spike-and-slab})
treats all candidate variables $X_i$ equally. However, in some
settings we may have additional information that suggests the
importance of some variables more than others.
\pkg{varbvs} can encode these preferences with a non-exchangeable
prior $\bm{\pi} = (\pi_1, \ldots, \pi_p)$, which is specified by
setting input \code{logodds} to a matrix with rows corresponding to
variables and columns corresponding to hyperparameter settings. We
demonstrate a non-exchangeable prior in one of the examples below.

An alternative to specifying a grid of hyperparameter settings 
is to estimate one or more of the hyperparameters.
This option is activated by setting \code{update.sigma = TRUE} and/or
\code{update.sa = TRUE} in \code{varbvs}, and it is activated by
default when \code{sigma} or \code{sa} are not specified. For
estimating one or more of the hyperparameters, we implemented a fast
approximate expectation maximization (EM) approach \citep{heskes-2004,
  neal-1998} in which the E-step is approximated using the
variational techniques described below.
%


The $Z_i$'s are additional predictors that are always included in the
model. Note that an intercept ($Z_i = 1$) is always included so the
user should never provide and intercept as one of the covariates. The
$Z_i$'s are assigned an improper, uniform prior ({\em i.e.}, a normal
prior with large variance). This prior is convenient because the
covariates are easily integrated out from the linear model
\citep{chipman-2001}, as well as the logistic regression model after
introducing an additional variational approximation (see the
Appendix). We caution that improper priors can result in improper
posteriors and Bayes factors \citep{obrien-2004}.

\subsection{Fast posterior computation via variational approximation}
\label{sec:posterior}

We use an alternative to MCMC \citep{george-1993} based on variational
methods \citep{blei-2016, jordan-1999, ormerod-2010, wainwright-2008}
that yields fast computation of posterior probabilities at the cost of
sometimes requiring a more careful interpretation due to the
approximations made. The basic idea is to recast the problem of
computing posterior probabilities---which is inherently an
intractable, high-dimensional integration problem---as an optimization
problem. This is achieved by introducing a class of approximating
distributions, then optimizing a criterion (the Kullback-Leibler
divergence) to find the distribution within this class that best
matches the posterior. To make this approach viable for large
problems, we enforce a simple conditional independence approximation
\citep{carbonetto-2012, logsdon-2010}: conditioned on the
hyperparameters $\theta \equiv \{\sigma^2$, $\sigma_a^2$,
$\bm{\pi}\}$, each regression coefficient $\beta_i$ is independent of
the other regression coefficients {\em a posteriori}. We then search
for a distribution with this conditional independence property that
best ``fits'' the posterior.
This conditional independence assumption was initially motivated from
the GWAS setting in which the variables are genetic markers. For more
details, see \cite{carbonetto-2012}.


The algorithm for fitting the variational approximation consists of an
inner loop and an outer loop. The outer loop iterates over the
hyperparameter grid points, and is described in the next section
(Sec.~\ref{sec:posterior-hyperparameters}). The inner loop, given a
setting of the hyperparameters, cycles through co-ordinate ascent
updates that try to minimize the Kullback-Leibler divergence between
the approximate posterior and exact posterior.
The inner loop co-ordinate ascent updates terminate when either the
maximum number of inner loop iterations is reached, as specified by
input \code{maxiter}, or the maximum difference between the estimated
posterior inclusion probabilities is less than \code{tol}. The
computational complexity of the co-ordinate ascent updates scales
linearly with the number of variables and the number of samples. The
number of co-ordinate ascent updates required to reach convergence
depends on the covariance structure of the candidate variables;
fastest convergence occurs when the variables are uncorrelated or
weakly correlated.

Function \code{varbvs} outputs three posterior quantities for each
variable $X_i$ and for each hyperparameter setting
$\theta^{(j)}$: 
\begin{align}
\alpha_{ij} &\approx \Pr(\beta_i \neq 0 \,|\, {\bf X}, {\bf Z},
                     \theta = \theta^{(j)}) \label{eq:alpha} \\
\mu_{ij} &\approx \mathrm{E}[\beta_i\,|\,{\bf X}, {\bf Z},
          \theta = \theta^{(j)}, \beta_i \neq 0] \label{eq:mu} \\
s_{ij}^2 &\approx \mathrm{Var}[\beta_i\,|\,{\bf X}, {\bf Z},
          \theta = \theta^{(j)}, \beta_i \neq 0]. \label{eq:s}
\end{align}
Each of these outputs is represented as a $p \times n_s$ matrix, where
$p$ is the number of variables and $n_s$ is the number of
hyperparameter grid points. For the $i$th variable and $j$th
hyperparameter setting, \code{alpha[i,j]} is the variational estimate
of the PIP (Equation~\ref{eq:alpha}), \code{mu[i,j]} is the
variational estimate of the posterior mean coefficient given that it
is included in the regression model (Equation~\ref{eq:mu}), and
\code{s[i,j]} is the estimated posterior variance
(Equation~\ref{eq:s}). Many other posterior statistics can be easily
derived from these outputs. For example, \code{alpha * mu} gives the
marginal posterior mean estimates of the regression coefficients.

%

These posterior statistics are also the free parameters of the
approximating distribution; that is, they are the parameters that are
optimized as part of the ``inner loop.'' An additional set of free
parameters is needed for the logistic regression model, and the fitted
values for these parameters are returned as $n \times n_s$ matrix
\code{eta}. When a good guess of the variational parameters are
available in advance, they can be used to initialize the co-ordinate
ascent algorithm by specifying inputs \code{alpha}, \code{mu}, \code{s}
and \code{eta} to function \code{varbvs}.


\subsection{Averaging over the hyperparameters}
\label{sec:posterior-hyperparameters}

In the simplest case, the hyperparameter vector $\theta = (\sigma^2,
\sigma_a^2, \bm{\pi})$ is known, or fixed, and \pkg{varbvs} can fit
the model and compute approximate posteriors $\Pr(\beta \,|\, {\bf X},
{\bf Z}, {\bf y}, \theta)$. The variational method also yields an
approximation (actually, a lower bound) to the marginal likelihood
$\Pr(y \,|\, {\bf X}, {\bf Z}, \theta)$ integrating over the
coefficients $\beta$. We denote this lower bound by
$\mathsf{LB}(\theta)$. This scheme is conceptually simple, but the
results may be sensitive to choice of the hyperparameters $\theta$. It
is analogous to fixing $\lambda$ in \pkg{glmnet} rather than
estimating it by cross-validation.

A natural alternative is to estimate $\theta$. The simplest way to do
this is to treat $\mathsf{LB}(\theta)$ as if it were the likelihood
and maximize $\mathsf{LB}(\theta)$ over $\theta$; that is, compute
$\hat{\theta} = \mathrm{argmax}_{\theta} \; \mathsf{LB}(\theta)$ and
report approximate posteriors $\Pr(\beta \,|\, {\bf X}, {\bf Z}, {\bf
  y}, \hat{\theta})$. This is analogous to estimating $\lambda$ in
\pkg{glmnet} by cross-validation. This is usually preferable to fixing
$\theta$ by hand, and it has the practical advantage of having little
computational overhead and does not require the user to specify a
prior. But it does not take account of uncertainty in the
hyperparameters, nor does it allow for incorporation of prior
information about the hyperparameters.

To address these limitations, we can introduce a prior on the
hyperparameters. The \pkg{varbvs} package allows any discrete uniform
prior on $\theta$: just specify a grid of values $\theta_1, \dots,
\theta_{n_s}$ and it will treat the prior on the hyperparameters as
uniform on that grid. It will use the lower bound to the likelihood to
approximate the posterior on $\theta$, so $\Pr(\theta = \theta^{(j)}
\,|\, {\bf X}, {\bf Z}, y)$ is approximated by $ w^{(j)} =
\mathsf{LB}(\theta^{})/\sum_{j'=1}^{n_s} \mathsf{LB}(\theta^{(j')})$.
Further, it implements the {\em Bayesian model averaging}
\citep{hoeting-1999}, computing approximate posteriors on $\beta$ by
averaging over this approximate posterior; {\em i.e.}, $\Pr(\beta
\,|\, {\bf X}, {\bf Z}, y) \approx \sum_{j=1}^{n_s} w^{(j)} \Pr(\beta
\,|\, {\bf X}, {\bf Z}, y, \theta^{(j)})$. This has the advantage of
incorporating prior information and taking account of uncertainty
along with a manageable increase in computational cost. We have made
the model averaging approach the recommended option in \pkg{varbvs},
although it does require the user to specify the prior, which may be
off-putting to some people. (To make this less painful, we provide
guidelines in the package documentation. For example, we recommend
setting the prior on $\sigma_a^2$ indirectly through the proportion of
variance in $y$ explained by $\bf X$; see \cite{guan-2011,
  zhou-2012}.)

We have also implemented a hybrid approach, which allows the user to
specify a prior on some of the hyperparameters while maximizing the
others. In fact, the default in \pkg{varbvs} is to estimate $\sigma^2$
and $\sigma_a^2$, and assign an exchangeable prior for $\pi$ that is
uniform on the log-odds scale. This was the approach used in the
leukemia example above, in which posterior probabilities were
approximated at 20 grid points of $\pi$ ranging from $10^{-3.5}$ to
$10^{-1.0}$ (Fig.~\ref{fig:varbvs-leukemia}, right-hand
plot). Importantly, this hybrid approach provides flexibility for
tackling large data sets, and it is used in most of the larger-scale
examples below.

One practical issue with the variational computation strategy is that
the variational approximation can be sensitive to the choice of
starting point $\theta^{\sf (init)}$.
To provide a more accurate variational approximation of the posterior
distribution, the optimization procedure is run in two stages by
default. In the first stage, the entire procedure is run to
completion, then the fitted variational parameters (stored in outputs
\code{alpha}, \code{mu}, \code{s}, \code{eta}) corresponding to the
maximum marginal likelihood are used to initialize the co-ordinate
ascent updates in the second stage.
The final posterior estimates tend to be more accurate using this
two-stage optimization approach \citep{carbonetto-2012}. Set
\code{initialize.params = FALSE} in \code{varbvs} to skip over the
initialization phase.

\subsection[The varbvs function]{The \code{varbvs} function}
\label{sec:interface}

We end this section with an overview of the core package function for
all BVS posterior computation and model fitting procedures in the
\proglang{R} package. To provide a familiar interface, we have modeled
it  after \pkg{glmnet}. The inputs to \code{varbvs} are grouped by
their function:
\begin{Code}
varbvs(X, Z, y, family,                              # Data.
       sigma, sa, logodds,                           # Hyperparameter grid.
       alpha, mu, eta,                               # Variational parameters.
       update.sigma, update.sa, optimize.eta,        # Optimization and model
       initialize.params, nr, sa0, n0, tol, maxiter, # fitting settings.
       verbose)                                      # Other settings.
\end{Code}
The first four input arguments are for the data: the $n \times p$
input matrix \code{X} and the $n \times m$ input matrix \code{Z},
where $n$ is the number of data examples, $p$ is the number of
candidate variables and $m$ is the number of covariates (not including
the intercept); the $n$ observations of the regression outcome,
\code{y}; and the option to specify a linear regression model
(\code{family = "gaussian"}, the default) or logistic regression when
all entries of \code{y} are 0 or 1 (\code{family = "binomial"}). 

The next three input arguments, \code{sigma}, \code{sa} and
\code{logodds}, are optional, and specify the grid of hyperparmeter
settings. Each of these inputs must be a single value, or have the
same number of entries $n_s$, except in the special case when the
prior inclusion probability is specified separately for each variable,
in which case \code{logodds} is a $p \times n_s$ matrix. If inputs
\code{sigma} or \code{sa} are missing, they are automatically fitted
to the data by computing approximate maximum-likelihood or {\em
  maximum a posteriori} estimates.

When good initial estimates of the variational parameters are
available, they can be provided to \code{varbvs} through input
arguments \code{alpha}, \code{mu} and \code{s}. Each of these inputs
must be an $p \times n_s$ matrix, or a $p \times 1$ matrix when all
variational approximations are provided the same initial parameter
estimate. Input \code{eta} is an additional set of free parameters for
the variational approximation to the logistic regression model. It is
either an $n \times n_s$ matrix or an $n \times 1$ matrix. The
remaining input arguments control various aspects of the model fitting
and optimization procedures, and are detailed in the \code{varbvs}
help page.

The \code{varbvs} function returns an S3 object of class \code{"varbvs"}.
The main components of interest are:
\begin{itemize}

  \item \code{logw}---Array in which \code{logw[i]} is the variational
    approximation to the marginal log-likelihood for the $i$th
    hyperparameter grid point.
    
  \item \code{w}---Approximate posterior probabilities, or ``weights,''
    $w^{(j)}$ computed from \code{logw}.
    
  \item \code{alpha}---Variational estimates of posterior inclusion
    probabilities, $\alpha_{ij}$, for each variable $X_i$ and
    hyperparameter setting $\theta^{(j)}$.

  \item \code{mu}---Variational estimates of posterior mean
    coefficients, $\mu_{ij}$, for each variable $X_i$ and hyperparameter
    setting $\theta^{(j)}$.

  \item \code{s}---Variational estimates of posterior variances,
    $s_{ij}$, for each variable $X_i$ and hyperparameter setting
    $\theta^{(j)}$.

  \item \code{pip}---The ``averaged'' posterior inclusion
    probabilities computed as a weighted sum of the individual PIPs
    (\code{alpha}), with weights given by \code{w}.
    
  \item \code{mu.cov}---Posterior mean regression coefficients
    \code{mu.cov[i,j]} for each covariate $Z_i$ (including the
    intercept) for each hyperparameter setting $\theta^{(j)}$.
  
  \item \code{eta}---Additional variational parameters for
    \code{family = "binomial"} only.
  
  \item \code{pve}---For each hyperparameter setting $\theta^{(j)}$,
    and for each variable $X_i$, \code{pve[i,j]} is the mean estimate
    of the proportion of variance in the outcome $Y$ explained by
    $X_i$, conditioned on $X_i$ being included in the model. This is
    computed for \code{family = "gaussian"} only.
    
  \item \code{model.pve}---Samples drawn from the posterior
    distribution giving estimates of the proportion of variance in the
    outcome $Y$ explained by the fitted variable selection model. For
    example, \code{mean(fit.varbvs$model.pve)} yields the posterior
    mean of the proportion of variance explained, where
    \code{fit.varbvs} is the \code{varbvs} return value. This is
    provided for \code{family = "gaussian"} only.

\end{itemize}

The components \code{alpha}, \code{mu}, \code{s} and \code{w} are
basic posterior quantities that can be used to quickly calculate many
other posterior statistics of interest.
For example, the probability that at least 1 variable is included in
the regression model is computed as
\begin{Code}
R> p0 <- apply(1 - fit$alpha, 2, prod)
R> sum(fit$w * (1 - p0)))
\end{Code}

The \pkg{varbvs} \proglang{R} package also provides standard
supporting functions for the \code{"varbvs"} class, including
\code{summary}, \code{predict} and \code{plot}.

\section{Example: mapping a complex trait in outbred mice} 
\label{sec:cfw}

In our second example, we illustrate the features of \pkg{varbvs} for
genome-wide mapping of a complex trait. The data, downloaded from
Zenodo \citep{zenodo}, are body and testis weight measurements
recorded for 993 outbred mice, and genotypes at 79,748 single
nucleotide polymorphisms (SNPs) for the same mice
\citep{parker-2016}. Our main aim is to identify genetic variants
contributing to variation in testis weight. The genotype data are
represented in \proglang{R} as a $993 \times 79,748$ matrix,
\code{geno}. The phenotype data---body and testis weight, in
grams---are stored in the \code{"sacwt"} and \code{"testis"} columns
of the \code{pheno} matrix:

\begin{Schunk}
\begin{Sinput}
R> head(pheno[, c("sacwt", "testis")])
\end{Sinput}
\begin{Soutput}
      sacwt testis
26305  46.6 0.1396
26306  35.7 0.1692
26307  34.1 0.1878
26308  41.8 0.2002
26309  39.5 0.1875
26310  36.0 0.1826
\end{Soutput}
\end{Schunk}

\begin{figure}[t!]
\begin{center}
\includegraphics[width=3.5in,keepaspectratio=true]{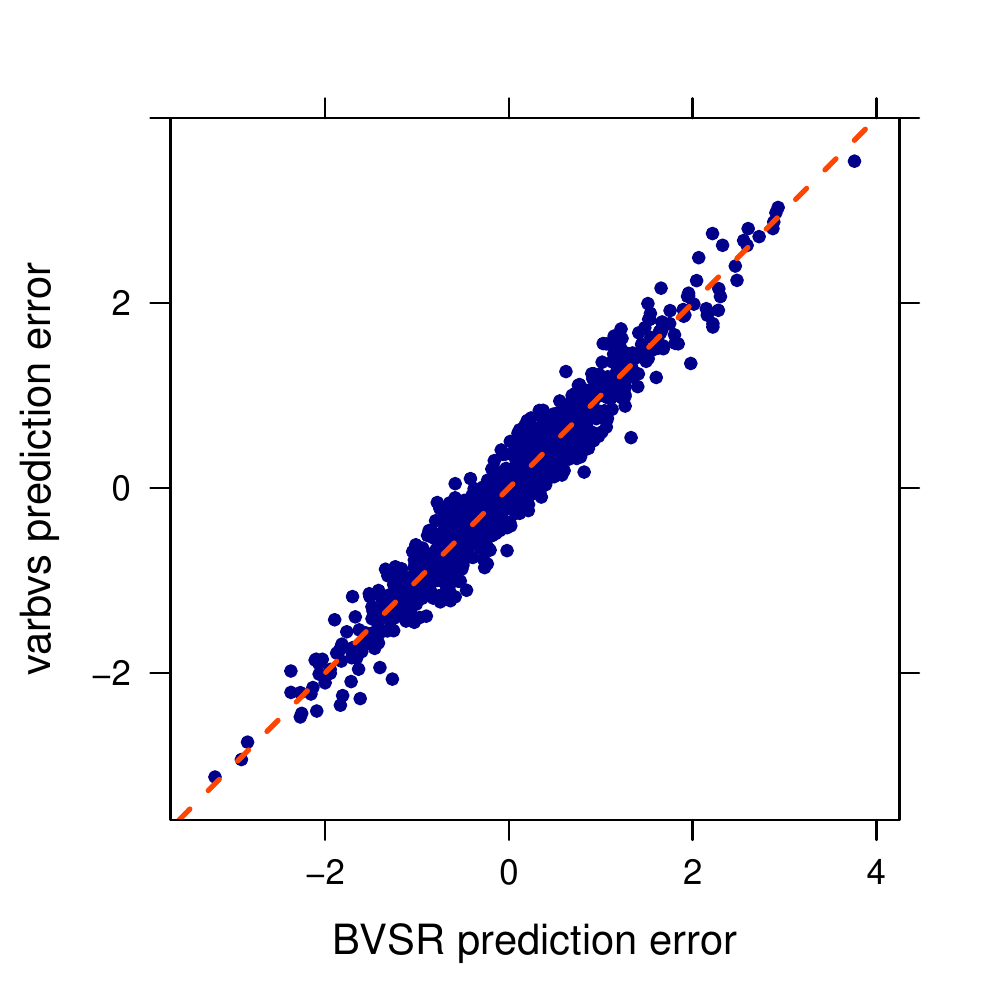}
\end{center}
\caption{Scatterplot comparing accuracy of \pkg{varbvs} and BVSR
  predictions. To assess prediction accuracy, we perform a simple
  cross-validation experiment in which the mouse data are split evenly
  into 10 test data sets: in each of the 10 rounds of
  cross-validation, the \pkg{varbvs} and BVSR models are fit to the
  remaining training samples, then the predictions are evaluated in
  the test set. The $x$ and $y$ axes in the plot show differences
  between predicted and observed phenotype (testis weight controlling
  for body weight) in the left-out test samples. These differences are
  normalized by the standard deviation of the phenotype computed from
  the full sample. The adjoining script \code{cfw.cv.R} reproduces the
  results of the cross-validation experiment, as well as this
  figure. The BVSR method is implemented in GEMMA version 0.96.}
\label{fig:cfw-cv} 
\end{figure}

The ``cfw'' vignette in the \proglang{R} package reproduces all the
results of this analysis except for Fig.~\ref{fig:cfw-cv}, which can
be reproduced by running script \code{cfw.cv.R} accompanying this
paper.

The standard approach in genome-wide mapping is to quantify support
for a quantitative trait locus (QTL) separately at each SNP. For
example, this was the approach taken in \cite{parker-2016}. Here, we
implement this univariate regression (``single-marker'') mapping
approach using the \code{-lm 2} option in GEMMA version 0.96
\citep{zhou-2012}, which returns a likelihood-ratio test {\em p}~value
for each SNP. We compare this single-marker analysis against a
\pkg{varbvs} multiple regression (``multi-marker'') analysis of the
same data.

In the \pkg{varbvs} analysis, the quantitative trait (testis weight)
is modeled as a linear combination of the covariate (body weight) and
the candidate variables (the 79,748 SNPs). As before, the model
fitting is accomplished with a single function call:

\begin{Schunk}
\begin{Sinput}
R> fit <- varbvs(geno, as.matrix(pheno[, "sacwt"]), pheno[, "testis"], 
+      sa = 0.05, logodds = seq(-5, -3, 0.25))
\end{Sinput}
\end{Schunk}

This call is completed in less than 4 minutes on a MacBook Air with a
1.86 GHz Intel CPU, 4 GB of memory and R 3.3.3. Note that, to simplify
this example, we have fixed \code{sa} to \code{0.05}, a choice
informed by our power calculations. In this application, it would be
preferable to average over a range of settings to avoid sensitivity to
prior choice.

Once the model fitting is completed, we quickly generate a summary of
the results using the \code{summary} function:

\begin{Schunk}
\begin{Sinput}
R> print(summary(fit))
\end{Sinput}
\begin{Soutput}
Summary of fitted Bayesian variable selection model:
family:     gaussian   num. hyperparameter settings: 9
samples:    993        iid variable selection prior: yes
variables:  79748      fit prior var. of coefs (sa): no
covariates: 2          fit residual var. (sigma):    yes
maximum log-likelihood lower bound: 2428.7093
proportion of variance explained: 0.149 [0.090,0.200]
Hyperparameters: 
        estimate Pr>0.95             candidate values
sigma   0.000389 [0.000379,0.000404] NA--NA
sa            NA [NA,NA]             0.05--0.05
logodds    -3.78 [-4.25,-3.50]       (-5.00)--(-3.00)
Selected variables by probability cutoff:
>0.10 >0.25 >0.50 >0.75 >0.90 >0.95 
    3     3     3     2     2     1 
Top 5 variables by inclusion probability:
            index    variable   prob    PVE     coef  Pr(coef.>0.95)
rs6279141   59249   rs6279141 1.0000 0.0631 -0.00806 [-0.010,-0.007]
rs33217671  24952  rs33217671 0.9351 0.0220  0.00509 [+0.003,+0.007]
rs33199318   9203  rs33199318 0.6869 0.0170  0.00666 [+0.004,+0.009]
rs52004293  67415  rs52004293 0.0739 0.0136  0.00347 [+0.002,+0.005]
rs253722776 44315 rs253722776 0.0707 0.0133 -0.00369 [-0.005,-0.002]
\end{Soutput}
\end{Schunk}

This summary tells us that only 3 out of the 79,748 SNPs are included
in the model with posterior probability greater than 0.5, and that the
included SNPs explain 15\% of the variance of testis weight.
(Precisely, this is the variance explained in testis weight residuals
after controlling for body weight.) Further, a single SNP (rs6279141)
accounts for over 6\% of variance in testis weight. This SNP is
located on chromosome 13 approximately 1 Mb from {\sf\em Inhba}, a
gene that has been previously shown to affect testis morphogenesis
\citep{mendis-2011, mithraprabhu-2010, tomaszewski-2007}.

\begin{figure}[t!]
\begin{center}
\includegraphics{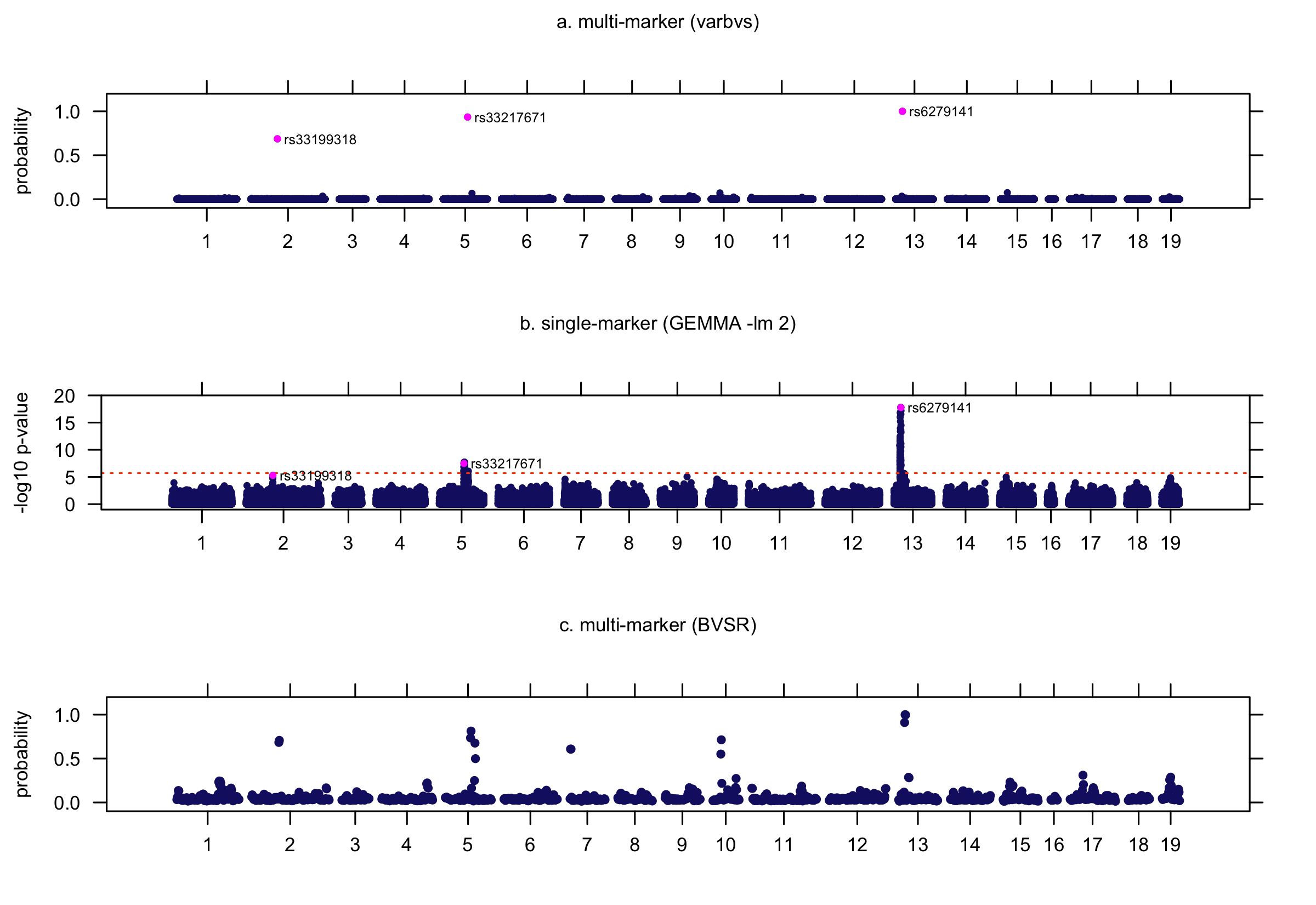}
\end{center}
\caption{QTL mapping of a complex trait in outbred mice. (a) Posterior
  inclusion probabilities for all 79,748 candidate SNPs on chromosomes
  1--19 computed using \code{varbvs}.
  SNPs with PIPs greater than 0.5 are highlighted. (b) {\em p}~values
  for the same candidate SNPs computed using GEMMA.
  threshold determined via permutation analysis, at $\mbox{{\em
      p}~value} = 2 \times 10^{-6}$ \citep{parker-2016}. (c) Posterior
  probabilities computed using the BVSR method in GEMMA version 0.96
  \citep{zhou-2013}. In BVSR, since multiple correlated SNPs at a
  single QTL are expected to be included in the model with lower
  probability, plotting individual PIPs does not highlight the
  QTLs. Therefore, the results are summarized by dividing each
  chromosome into contiguous segments containing 100 SNPs, in which
  consecutive segments overlap by 50 SNPs, and computing the posterior
  probability that at least 1 SNP is included within each of these
  segments.}
\label{fig:cfw} 
\end{figure}

We can also quickly create a visual summary of the results using the
\code{plot} function:

\begin{Schunk}
\begin{Sinput}
R> print(plot(fit, vars = c("rs33199318", "rs33217671", "rs6279141"), 
+      groups = map$chr, gap = 1500))
\end{Sinput}
\end{Schunk}

The output is shown in Fig.~\ref{fig:cfw}a. Note that the \code{plot}
function has a \code{"group"} argument, which allows us to arrange the
variable selection results by chromosome.


It is informative to compare these probabilities against the
``single-marker'' {\em p}~values that ignore correlations between SNPs
(Fig.~\ref{fig:cfw}b). 
Reassuringly, the loci with the strongest support for association in
the single-marker analysis (Fig.~\ref{fig:cfw}b) also exhibit the
strongest support for association in the multi-marker analysis
(Fig.~\ref{fig:cfw}a). Further, SNPs included with the highest
posterior probabilities are among the SNPs with the smallest {\em
  p}~values. One QTL on chromosome 2 is not significant in the
single-marker analysis ($\mbox{{\em p}~value} = 5.2 \times 10^{-6}$),
yet shows moderate probability of association in the multi-marker
analysis. The multi-marker association signal at this locus is
concentrated in a small region that contains a single gene, {\sf\em
  Myo3b}, providing an additional testis weight gene for further
investigation.

In Fig.~\ref{fig:cfw}b, we observe that many SNPs have low {\em
  p}~values at each of the identified testis weight loci. This
illustrates the common situation in GWAS in which many SNPs at a
single locus are associated with the trait. In general, when multiple
variables are strongly correlated with each other, the
fully-factorized variational approximation in \pkg{varbvs} tends to
concentrate the posterior mass on a single variable.
%
%

We also assessed the accuracy of the variational approximation by
comparing the \pkg{varbvs} results (Fig.~\ref{fig:cfw}a) against
another method, BVSR \citep{zhou-2013}, that uses a very similar
model. The BVSR method implemented in GEMMA uses MCMC to estimate
posterior probabilities. Comparing panels a and c in
Fig.~\ref{fig:cfw}, BVSR yields a more complex model in which hundreds
of SNPs are included in the model with low probability---yet the loci
with the strongest support in the \pkg{varbvs} and BVSR analyses
closely agree. The simpler \pkg{varbvs} model also achieves similar
prediction accuracy to the BVSR model; in a simple cross-validation
experiment in which 10\% of the samples in each round are used to test
the model, the prediction errors of the \pkg{varbvs} and BVSR methods
are 97\% correlated (Fig.~\ref{fig:cfw-cv}).


\section{Example: mapping Crohn's disease risk loci}
\label{sec:cd}

Our third example again illustrates \pkg{varbvs}'s ability to tackle
large data sets for mapping genetic loci contributing to a complex
trait. The data set in this example contains 4,686 samples (1,748
Crohn's disease cases, 2,938 controls) and 442,001 SNPs
\citep{wtccc-2007}. The genotypes are stored in a $4,686 \times
442,001$ matrix \code{X}, and the binary outcome is disease status (0
= control, 1 = case):
\begin{CodeChunk}
\begin{CodeInput}
> print(summary(factor(y)))
\end{CodeInput}
\begin{CodeOutput}
   0    1
2938 1748
\end{CodeOutput}
\end{CodeChunk}
We model Crohn's disease disease status using logistic regression,
with the 442,001 SNPs as candidate variables, and no additional
covariates. On a machine with a 2.5 GHz Intel Xeon CPU, fitting the
BVS model to the data took 39 hours to complete:
\begin{Code}
R> fit <- varbvs(X, NULL, y, family = "binomial", 
                 logodds = seq(-6,-3,0.25), n0 = 0)
\end{Code}
The ``cd'' vignette reproduces all the results and plots shown here.
Since the data needed to run the script cannot be made publicly
available due to data sharing restrictions, those wishing to reproduce
this analysis must apply for data access by contacting the Wellcome
Trust Case Control Consortium. 

Similar to the previous examples, the fitted regression model is very
sparse; only 8 out of the 442,001 candidate variables are included in
the model with probability at least 0.5:
\begin{CodeChunk}
\begin{CodeInput}
R> print(summary(fit,nv = 9))
\end{CodeInput}
\begin{CodeOutput}
Summary of fitted Bayesian variable selection model:
family:     binomial   num. hyperparameter settings: 13
samples:    4686       iid variable selection prior: yes
variables:  442001     fit prior var. of coefs (sa): yes
fit approx. factors (eta):    yes
maximum log-likelihood lower bound: -3043.2388
Hyperparameters:
        estimate Pr>0.95             candidate values
sa         0.032 [0.0201,0.04]       NA--NA
logodds    -4.06 [-4.25,-3.75]       (-6.00)--(-3.00)
Selected variables by probability cutoff:
>0.10 >0.25 >0.50 >0.75 >0.90 >0.95
   13    10     8     7     7     7
Top 9 variables by inclusion probability:
     index   variable  prob PVE  coef*  Pr(coef.>0.95)
  1  71850 rs10210302 1.000  NA -0.313 [-0.397,-0.236]
  2  10067 rs11805303 1.000  NA  0.291 [+0.207,+0.377]
  3 140044 rs17234657 1.000  NA  0.370 [+0.255,+0.484]
  4 381590 rs17221417 1.000  NA  0.279 [+0.192,+0.371]
  5 402183  rs2542151 0.992  NA  0.290 [+0.186,+0.392]
  6 271787 rs10995271 0.987  NA  0.236 [+0.151,+0.323]
  7 278438  rs7095491 0.969  NA  0.222 [+0.141,+0.303]
  8 168677  rs9469220 0.586  NA -0.194 [-0.269,-0.118]
  9  22989 rs12035082 0.485  NA  0.195 [+0.111,+0.277]
*See help(varbvs) about interpreting coefficients in logistic regression.
\end{CodeOutput}
\end{CodeChunk}

\begin{figure}[t!]
\includegraphics[width=6.2in,keepaspectratio=true]{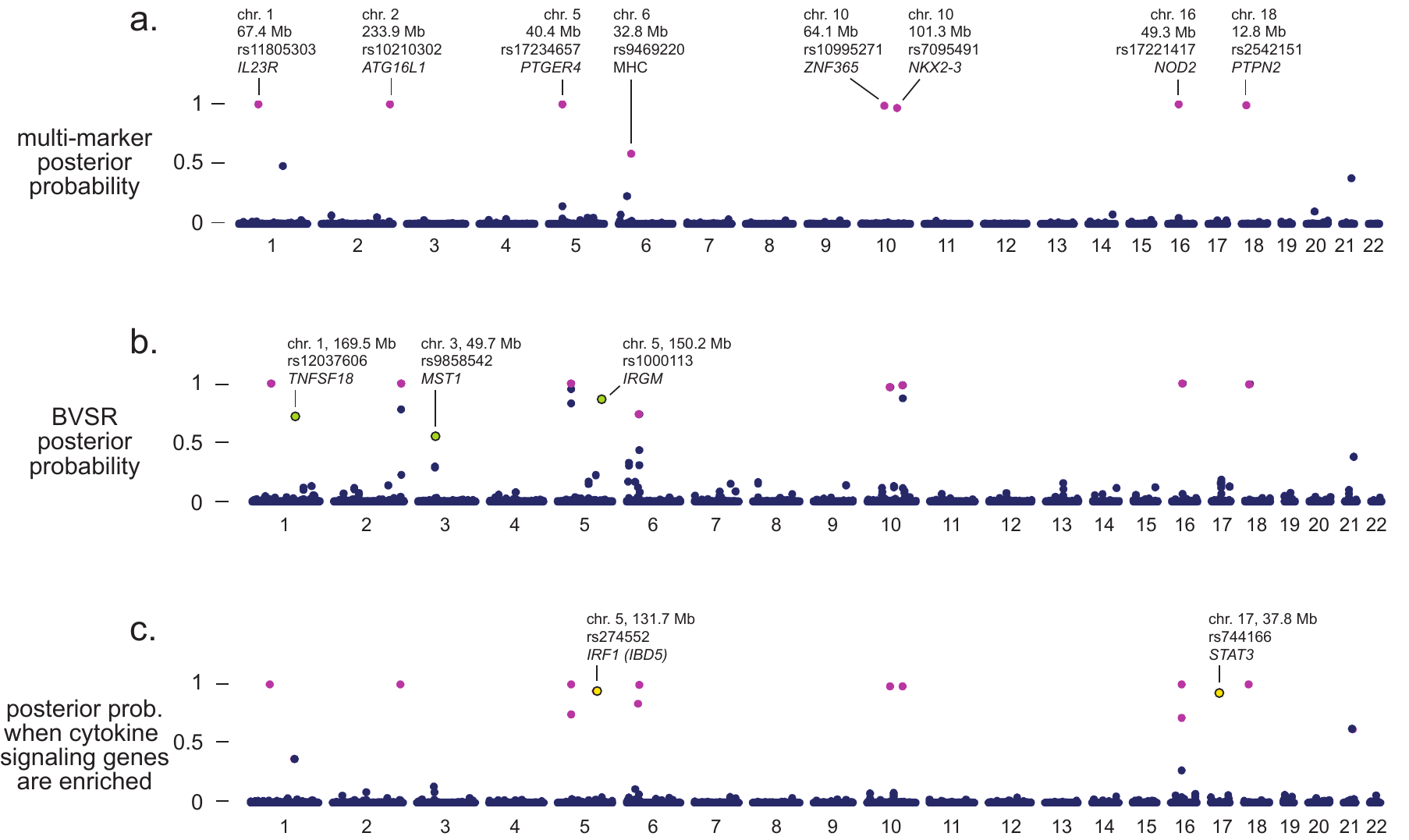}
\caption{\pkg{varbvs} and BVSR analysis of Crohn's disease data. (a)
  Posterior inclusion probabilities for all 442,001 candidate SNPs on
  chromosomes 1--22. SNPs with PIP greater than 0.5 are highlighted. 
  Human Genome Assembly hg17 (NCBI release 35). (b) Posterior
  probabilities estimated in BVSR \citep{zhou-2013}. Similar to the
  mouse data, each chromosome is divided into overlapping 50-SNP
  segments, and the plot shows the posterior probability that at least
  1 SNP is included within each segment. Three points are highlighted
  in light green; these are segments with posterior probability
  greater than 0.5 in the BVSR analysis that do not contain a SNP with
  PIP greater than 0.5 in the \pkg{varbvs} analysis. (c) PIPs for all
  SNPs conditioned on enrichment of cytokine signaling genes. Two SNPs
  are highlights in yellow; they are the two SNPs with a PIP greater
  than 0.5 only after prioritizing SNPs near cytokine signaling genes.}
\label{fig:cd} 
\end{figure}

The \pkg{varbvs} results, summarized in Fig.~\ref{fig:cd}a, provide
strong support for nearly the same reported {\em p}~values at the
previously used ``whole-genome'' significance threshold, $5 \times
10^{-7}$; in particular, the 7 SNPs included in the regression model
with probability greater than 0.9 correspond to the smallest trend
{\em p}~values, between $7.1 \times 10^{-14}$ and $2.68 \times
10^{-7}$ \citep{wtccc-2007}. Additionally, the SNP the highest
posterior probability is most cases the exact same SNP with the
smallest trend {\em p}~value. (See \citealt{carbonetto-2013} for an
extended comparison of the {\em p}~values and PIPs.) Only one disease
locus, near gene {\sf\em IRGM} on chromosome 5, has substantially
stronger support in the single-marker analysis; the originally
reported {\em p} value is $5.1 \times 10^{-8}$, whereas the
\pkg{varbvs} analysis yields a largest posterior probability of $0.05$
at this locus.

To further validate the \pkg{varbvs} analysis of the Crohn's disease
data, we compared the \pkg{varbvs} results against posterior
probabilities computed using the BVSR method. As before, we obtain
similar variable selection results; the loci with the strongest
support in the \pkg{varbvs} analysis (Fig.~\ref{fig:cd}a) are the same
loci identified by the BVSR method (Fig.~\ref{fig:cd}b) aside from a
few loci with moderate support in the BVSR analysis near genes {\sf\em
  TNFSF18}, {\sf\em MST1} and {\sf\em IRGM}.

\section{Example: gene set enrichment analysis in Crohn's disease}
\label{sec:cytokine}

In this section, we revisit the Crohn's disease data set to
demonstrate the use of \pkg{varbvs} for model comparison. This
analysis is implemented in the ``cytokine'' vignette.

Here, we incorporate additional information about the
442,001 candidate variables, stored in a vector, \code{cytokine}:
\begin{CodeChunk}
\begin{CodeInput}
R> data(cytokine)
R> print(summary(factor(cytokine)))
\end{CodeInput}
\begin{CodeOutput}
     0      1
435290   6711
\end{CodeOutput}
\end{CodeChunk}

An entry of 1 means that the SNP is located within 100 kb of a gene in
the ``Cytokine signaling in immune system'' gene set. This gene set
was previously identified in an interrogation of 3,158 gene sets from
8 publicly available biological pathway databases
\citep{carbonetto-2013}.

To assess relevance of cytokine signaling genes to Crohn's disease
risk, we modify the prior so that SNPs near cytokine signaling genes
are included in the model with higher probability ({\em i.e.,}
cytokine signaling genes are ``enriched'' for Crohn's disease risk
loci). To simplify this example, the default prior log-odds is set to
\code{-4}, which is the maximum-likelihood value from the above
analysis. We evaluate 13 settings of the modified prior, ranging from
\code{-4} (1 out of 10,000 SNPs is included in the model) to \code{-1}
(approximately 1 out of 10 SNPs is included):
\begin{Code}
R> logodds <- matrix(-4,442001,13)
R> logodds[cytokine == 1,] <- matrix(seq(0,3,0.25) - 4,6711,13,byrow = TRUE)
\end{Code}
We then fit the BVS model to the data using this modified prior:
\begin{Code}
R> fit.cytokine <- varbvs(X, NULL, y, family = "binomial", 
                          logodds = logodds, n0 = 0)
\end{Code}

The new variable selection results are summarized in
Fig.~\ref{fig:cd}c. The SNPs identified in the previous analysis are
retained under the new prior. Further, 2 new SNPs, near genes {\sf\em
  IRF1} and {\sf\em STAT3}, show strong support for association only
after allowing for enrichment of associations near cytokine signaling
genes.

To assess support for this model, we compute a Bayes factor
\citep{kass-1995} that compares against the ``null'' model in which
all SNPs are equally likely to be included {\em a priori} ({\em i.e.},
an exchangeable prior):

\begin{CodeChunk}
\begin{CodeInput}
R> fit.null <- varbvs(X, NULL, y, "binomial", logodds = -4, n0 = 0)
R> BF <- varbvsbf(fit.null, fit.cytokine)
R> print(format(, scientific = TRUE))
\end{CodeInput}
\begin{CodeOutput}
[1] "9.355e+05"
\end{CodeOutput}
\end{CodeChunk}
This Bayes factor is strong evidence that Crohn's disease risk loci
are found with greater frequency near cytokine signaling genes.

\section{Summary and discussion}
\label{sec:discussion}

In this paper, we illustrated the benefits of Bayesian variable
selection techniques for regression analysis, and showed that
\pkg{varbvs} provides a user-friendly interface for applying BVS to
large data sets.
Mathematical details and derivations of the algorithms are found in
the Appendix and in \cite{carbonetto-2012}. In the remainder, we
provide some additional background and guidance.

As our examples illustrate, one benefit of BVS is that it provides a
measure of uncertainty in the parameter estimates.
Assessing uncertainty is often not done in practice because it
requires careful selection of priors. Therefore, we have provided
default priors that are suitable in many settings. This allows the
practitioner to expedite the analysis, and perhaps revisit the prior
choices at a later date. The default priors are based on detailed
discussions from our earlier work \citep{guan-2011, servin-2007,
  zhou-2013}.
As an alternative, \pkg{varbvs} also allows for computation of
hyperparameter point estimates.

Fast computation of posterior probabilities is made possible by the
formulation of a variational approximation derived from a simple
conditional independence assumption. Even when many of the variables
are strongly correlated, this approximation can often yield accurate
inferences so long as individual posterior statistics are interpreted
carefully. The computational complexity of the co-ordinate ascent
algorithm for fitting the variational approximation is linear in the
number of samples and in the number of variables so long as the
correlations between variables are mostly small. This makes the
algorithm suitable for many genetic data sets since correlations are
limited by recombination. However, for data sets with widespread
correlations between variables, convergence of the algorithm can be
slow. We are currently investigating faster alternatives using
quasi-Newton methods and acceleration schemes such as SQUAREM
\citep{varadhan-2008, squarem}.


In practice, final estimates can be sensitive to initialization of the
variational parameters. We have reduced this sensitivity by including
an additional optimization step that first identifies a good
initialization of the variational parameters
(Sec.~\ref{sec:posterior-hyperparameters}). However, it is good
practice to verify that different random initializations of these
parameters do not yield substantially different conclusions. The
documentation for function \code{varbvs} gives further guidance on
this, as well as guidelines for correctly interpreting variational
estimates of the posterior statistics.

\begin{figure}[t!]
\includegraphics[width=6.1in,keepaspectratio=true]{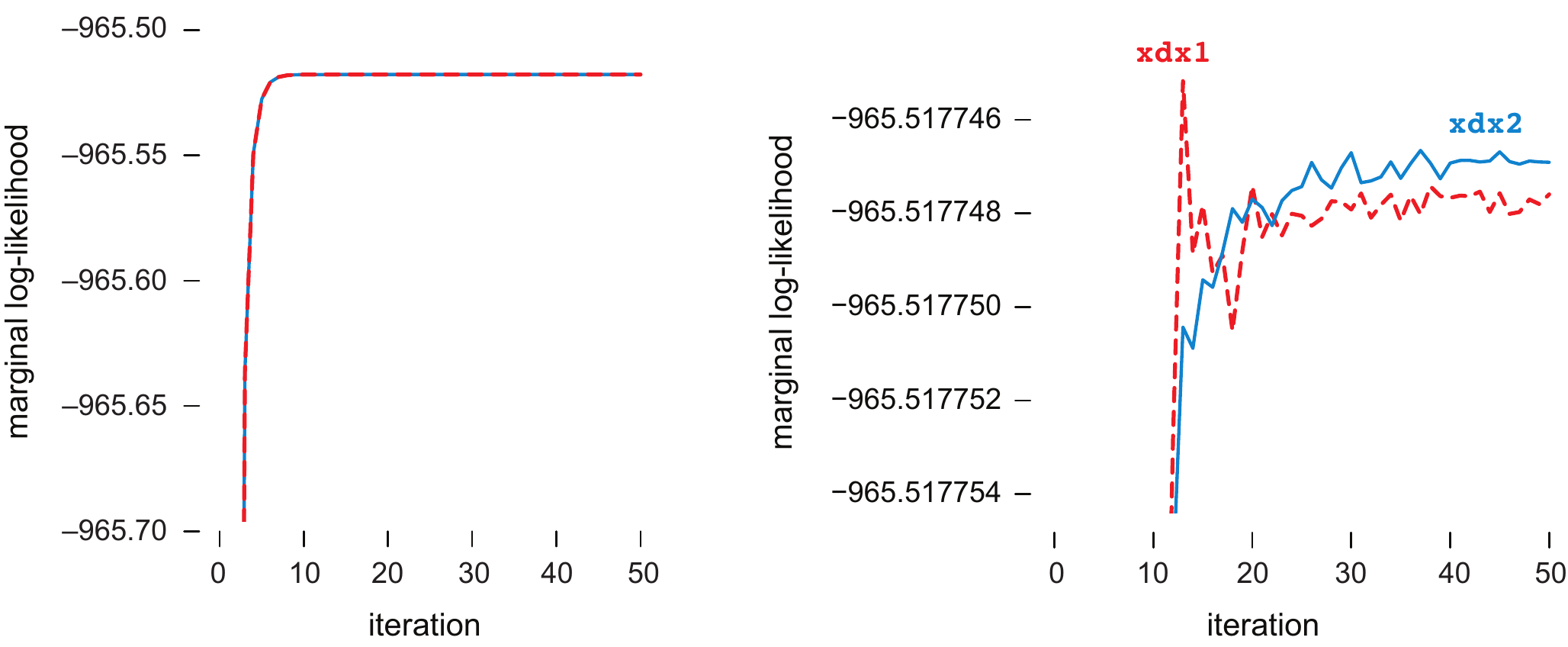}
\caption{Convergence of \pkg{varbvs} model fitting algorithm with less
  numerically stable (\code{xdx1}) and more numerically stable
  (\code{xdx2}) updates. The vertical axes show the variational lower
  bound to the marginal log-likelihood, which is also the objective
  function being maximized; the co-ordinate ascent updates terminate
  when they no longer increase the lower bound. The right-hand plot is
  a magnified version of the left-hand plot.}
\label{fig:stability} 
\end{figure}

Finally, we would like to remark on an often overlooked aspect of
statistical analyses---numerical stability.
In the logistic regression model, part of the variational optimization
algorithm involves computing the diagonal entries of the matrix
product ${\bf X}^T \hat{D} {\bf X}$, in which $\hat{D}$ is an $n
\times n$ diagonal matrix (see the Appendix). In the \proglang{MATLAB}
implementation, the following two lines of code are mathematically
equivalent,
\begin{Code}
xdx1 = diag(X'*D*X) - (X'*d).^2/sum(d)
xdx2 = diag(X'*D*X) - (X'*(d/sqrt(sum(d)))).^2
\end{Code}
where \code{d = diag(D)}. Yet, in floating-point arithmetic, the order
of operations affects the numerical precision of the final result,
which can in turn affect the stability of the co-ordinate ascent
updates. To illustrate this, we applied \pkg{varbvs}, using the two
different updates (\code{xdx1} and \code{xdx2}), to a data set with
simulated variables and a binary outcome. In Fig.~\ref{fig:stability},
we see that the second update (\code{xdx2}), corresponding to the
solid blue line in the plots, produced iterates that progressed more
smoothly to a stationary point of the objective function, whereas the
first update (\code{xdx1}) terminated prematurely because it produced
a large decrease in the objective. This illustrates the more general
point that numerical stability of operations can impact the quality of
the final solution, particularly for large data sets.

\section*{Acknowledgments}

Thanks to John Zekos and the Research Computing Center staff for their
support. Thanks to Abraham Palmer, Clarissa Parker, Shyam
Gopalakrishnan, Arimantas Lionikas, and other members of the Palmer
Lab for their contributions to the mouse data set. Thanks to Xiang
Zhu, Gao Wang, Wei Wang, David Gerard and other members of the
Stephens lab for their feedback on the code. Thanks to Cisca Wijmenga
and Gosia Trynka for their assistance with other genetic data analyses
that lead to important code improvements. Thank you to Karl Broman and
the authors of \pkg{glmnet} for providing excellent R packages that
have influenced the design of \pkg{varbvs}. Thanks to Ravi Varadhan
and Yu Du for their feedback, and thanks to Ann Carbonetto for her
support and encouragement.

\bibliography{varbvs-jss}

\appendix

\section{About this document}
\label{sec:session-info}

This manuscript was prepared using the \code{Sweave} function from the
\pkg{weaver} package \citep{weaver}. The code chunk below records the
version of R and the packages that were used to generate the results
contained in this manuscript.

\begin{Schunk}
\begin{Sinput}
R> sessionInfo()
\end{Sinput}
\begin{Soutput}
R version 3.4.1 (2017-06-30)
Platform: x86_64-apple-darwin15.6.0 (64-bit)
Running under: macOS Sierra 10.12.6

Matrix products: default
BLAS: /Library/Frameworks/R.framework/Versions/3.4/Resources/lib/libRblas.0.dylib
LAPACK: /Library/Frameworks/R.framework/Versions/3.4/Resources/lib/libRlapack.dylib

locale:
[1] en_US.UTF-8/en_US.UTF-8/en_US.UTF-8/C/en_US.UTF-8/en_US.UTF-8

attached base packages:
[1] methods   tools     stats     graphics  grDevices utils    
[7] datasets  base     

other attached packages:
 [1] varbvs_2.4-0        glmnet_2.0-10       foreach_1.4.3      
 [4] Matrix_1.2-10       latticeExtra_0.6-28 RColorBrewer_1.1-2 
 [7] lattice_0.20-35     curl_2.8.1          weaver_1.42.0      
[10] codetools_0.2-15    digest_0.6.12      

loaded via a namespace (and not attached):
[1] compiler_3.4.1  Rcpp_0.12.12    grid_3.4.1      iterators_1.0.8
\end{Soutput}
\end{Schunk}

\section{Additional derivations for linear regression model}
\label{sec:appendix-linear}

Most of the derivations for the linear regression model are given in
\cite{carbonetto-2012}. Here, we extend the variational approximation
to allow for additional variables $(Z_1, \ldots, Z_m)^T$ that are
included in the model with probability 1, and a non-exchangeable prior
on the regression coefficients $\beta_i$. Additionally, we derive an
approximate EM algorithm for the residual variance $\sigma^2$ and
prior variance $\sigma_a^2$.

First, we analytically integrate out the regression coefficients $u =
(u_1, \ldots, u_m)^T$ by making use of the following result:
\begin{equation}
|\Sigma_0|^{1/2} \Pr(y \,|\, {\bf X}, {\bf Z}, \beta, \sigma^2) =
|{\bf Z}^T{\bf Z}|^{-1/2} \Pr(\hat{y} \,|\,\hat {\bf X}, \beta, \sigma^2),
\end{equation}
in which $\Pr(y \,|\, {\bf X}, {\bf Z}, \beta, \sigma^2)$ is the
multivariate normal likelihood defined by the linear regression model
(Equation~\ref{eq:linear-regression}), $\Pr(\hat{y} \,|\,\hat {\bf X},
\beta, \sigma^2)$ is the likelihood given by linear regression
$\hat{y} = \hat{\bf X} \beta + \sigma^2$, $u$ is assigned a
multivariate normal prior with zero mean and covariance $\Sigma_0$
such that $|\Sigma_0^{-1}|$ is close to zero (yielding a ``flat''
prior density on $u$), and we define $\hat{\bf X} = {\bf X} - {\bf
  Z}({\bf Z}^T{\bf Z})^{-1} {\bf Z}^T {\bf X}$ and $\hat{y} = y - {\bf
  Z}({\bf Z}^T{\bf Z})^{-1} {\bf Z}^Ty$. Therefore, we can easily
account for the linear effects of covariates $Z$ by replacing all
instances of ${\bf X}$ with $\hat{\bf X}$ and all instances of $y$ with
$\hat{y}$, and by multipling the likelihood by $|{\bf Z}^T{\bf
  Z}|^{-1/2}$. Therefore, in the derivations below we assume the
simpler linear regression $y = {\bf X}\beta + \sigma^2$, replace $X$
with $\hat{X}$ and $y$ with $\hat{y}$, and multiply by $|{\bf Z}^T{\bf
  Z}|^{-1/2}$ to obtain the final solution.

The basic idea behind the variational approximation is to formulate a
lower bound to the marginal likelihood, $\Pr(y \,|\, \mathbf{X},
\theta) \geq \mathsf{LB}(\theta) \equiv e^{f({\bf X}, y, \theta,
  \phi)}$, then to adjust the free parameters, which we denote here by
$\phi \equiv \{\alpha, \mu, s\}$, so that this bound is as tight as
possible. This lower bound is formulated by introducing a probability
distribution $q(\beta; \phi)$ that approximates the posterior of
$\beta$ given $\theta$. Maximizing the lower bound corresponds to
finding the approximating distribution that best matches the
posterior; more precisely, it amounts to searching for the free
parameters $\phi$ that minimize the Kullback-Leibler divergence
between $q(\beta; \phi)$ and the posterior of $\beta$ given $\theta$
\citep{jordan-1999}.

The ``fully-factorized'' class of approximating distributions yields
the following analytical expression for the variational lower bound:
\begin{align}
f({\bf X}, y, \theta, \phi) &=
- \frac{n}{2}\log(2\pi\sigma^2)
- \frac{\|y - {\bf X}r\|_2^2}{2\sigma^2} 
- \frac{1}{2\sigma^2} \sum_{i=1}^p ({\bf X}^T{\bf X})_{ii}
  \mathrm{Var}[\beta_i] \nonumber \\
&\quad - \sum_{i=1}^p \alpha_i \log\Big(\frac{\alpha_i}{\pi_i}\Big)
       - \sum_{i=1}^p (1-\alpha_i) \log\Big(\frac{1-\alpha_i}{1-\pi_i}\Big)
  \nonumber \\
&\quad + \sum_{i=1}^p \frac{\alpha_i}{2} \bigg[
1 + \log\bigg(\frac{s_i^2}{\sigma_a^2 \sigma^2}\bigg)
  - \frac{s_i^2 + \mu_i^2}{\sigma_a^2 \sigma^2} \bigg],
\label{eq:lower-bound-expanded-linear}
\end{align}
where $\|\,\cdot\,\|_2$ is the Euclidean norm, $r$ is a column vector
with entries $r_i = \alpha_i \mu_i$, and $\mathrm{Var}[\beta_i] =
\alpha_i(s_i^2 + \mu_i^2) - (\alpha_i\mu_i)^2$ is the variance of
$i$th coefficient under the approximating distribution. As in
\cite{carbonetto-2012}, the co-ordinate updates for the free
parameters conditioned on a hyperparameter setting $\theta$ are
obtained by taking partial derivatives of the lower bound
(Equation~\ref{eq:lower-bound-expanded-linear}), setting these partial
derivatives to zero, and solving for the free parameters. This yields
the following expressions:
\begin{align}
\mu_i &= {\textstyle \frac{s_i^2}{\sigma^2}}
\big( ({\bf X}^Ty)_i - {\textstyle \sum_{j \neq i}}
({\bf X}^T{\bf X})_{ij} \alpha_j \mu_j \big)
\label{eq:update-mu} \\
s_i^2 &= \sigma^2/\big(({\bf X}^T{\bf X})_{ii} + 1/\sigma_a^2\big)
\label{eq:update-s} \\
\frac{\alpha_i}{1-\alpha_i} &= \frac{\pi_i}{1-\pi_i} \times 
\frac{s_i}{\sigma\sigma_a} \times e^{\mu_i^2/(2s_i^2)}.
\label{eq:update-alpha}
\end{align}

The E and M steps in the EM algorithm can be viewed as both minimizing
the Kullback-Leibler divergence \citep{neal-1998} or, equivalently in
this case, maximizing the lower bound
(Equation~\ref{eq:lower-bound-expanded-linear}). Therefore, we obtain
an ``approximate'' EM algorithm (e.g., \citealt{heskes-2004}) by
computing posterior expectations in the E-step under the assumption
that the true posterior is ``fully-factorized.'' We derive the M-step
updates for $\sigma^2$ and $\sigma_a^2$ in the standard way by solving
for roots $\sigma^2$ and $\sigma_a^2$ of the gradient, yielding
\begin{align}
\sigma^2 &= \frac{\|y - {\bf X}r\|_2^2 +
\sum_{i=1}^p ({\bf X}^T{\bf X})_{ii} \mathrm{Var}[\beta_i] 
+ \sum_{i=1}^p \alpha_i (s_i^2 + \mu_i^2)/\sigma_a^2)}
{n + \sum_{i=1}^p \alpha_i} \label{eq:update-sigma} \\
\sigma_a^2 &= \frac{\sum_{i=1}^p \alpha_i (s_i^2 + \mu_i^2)}
{\sigma^2 \sum_{i=1}^p \alpha_i}. \label{eq:update-sa}
\end{align}

\section{Additional derivations for logistic regression model}
\label{sec:appendix-logistic}

In the Appendix of \cite{carbonetto-2012}, we described an extension
to the fully-factorized variational approximation for Bayesian
variable selection with a logistic regression model and an
intercept. Here, we extend these derivations to allow for for
additional variables $Z = (Z_1, \ldots, Z_m)^T$ that are not subject
to the spike-and-slab priors.

We split the derivation into four parts: in the first part, we derive
a linear approximation to the non-linear likelihood; in the second
part, we analytically integrate out the coefficients $u$ from the
linearized likelihood; in the third part, we introduce the
fully-factorized variational approximation, and derive the co-ordinate
ascent updates for maximizing the variational lower bound; finally, in
the fourth part, we derive ``M-step'' updates for the additional free
parameters $\eta_i$ that were introduced to approximate the logistic
regression likelihood.

{\bf Taking care of the nonlinear factors in the likelihood.} For the
moment, we assume the simpler logistic regression with no additional
variables $Z$; it is easy to introduce these variables into the
expressions later on by substituting $\beta$ with $\binom{u}{\beta}$
and ${\bf X}$ with $({\bf Z\; X})$. The expression for the
log-likelihood given the simpler logistic regression can be written as
\begin{equation}
\log \Pr(y \,|\, {\bf X}, \beta) = 
(y-1)^T{\bf X}\beta + \sum_{i=1}^n \log p_i,
\label{eq:likelihood-logistic}
\end{equation}
in which we define $p_i \equiv \Pr(y_i = 1 \,|\, x_{i1}, \ldots,
x_{ip}, \beta) = \sigma(\sum_{j=1}^p x_{ij} \beta_j)$, and
$\sigma(x) = 1/(1 + e^{-x})$ is the sigmoid function (or inverse of
logit function). Written in this way, the linear components are
contained exclusively in the first term of
Equation~\ref{eq:likelihood-logistic}.

The basic idea behind the variational approximation is to formulate a
lower bound to the logarithm of the sigmoid function. Skipping the
technical details \citep{jaakkola-2000}, we obtain the following lower
bound:
\begin{equation}
\textstyle \log\sigma(x) \geq 
\log\sigma(\eta) + \frac{1}{2}(x - \eta) - \frac{d}{2}(x^2 - \eta^2),
\label{eq:sigmoid-lower-bound}
\end{equation}
in which we define $d = \frac{1}{\eta} (\sigma(\eta) - \frac{1}{2})$.
Notice that this expression introduces an additional parameter,
$\eta$. This identity holds for any choice of $\eta$, and this is the
free parameter that we will adjust to tighten the fit of the lower
bound as best as possible. We will have one free parameter $\eta_i$
for every factor in the likelihood. Also notice that all terms
involving $x$---later replaced by linear combinations of $\beta$---are
linear or quadratic in $x$.

Inserting this lower bound into the expression for the log-likelihood,
we obtain a lower bound to the log-likelihood, denoted by $g(\beta;
\eta)$:
\begin{equation}
g(\beta; \eta) = \sum_{i=1}^n \log\sigma(\eta_i) 
+ \textstyle \frac{\eta_i}{2}(d_i\eta_i - 1)
+ (y - \frac{1}{2})^T{\bf X}\beta -
\frac{1}{2}\beta^T{\bf X}^T\!D{\bf X}\beta,
\label{eq:lower-bound-logistic}
\end{equation}
where $D$ is the $n \times n$ matrix with diagonal entries $d_i$. By
extension, we have a lower bound on the marginal likelihood:
\begin{align}
\Pr(y \,|\, {\bf X}) &= 
\textstyle \int \Pr(y \,|\, {\bf X}, \beta) \, \Pr(\beta) \, d\beta 
\nonumber \\
&\geq \textstyle \int e^{g(\beta; \eta)} \, \Pr(\beta) \, d\beta.
\end{align}

{\bf Integrating out the coefficients $u$.} Since we have assigned an
(improper) normal prior to $u$ (with large variance), we can
analytically integrate out $u$ from the lower bound
(Equation~\ref{eq:lower-bound-logistic}), in which we substitute
$\beta$ with $\binom{u}{\beta}$, and we substitute ${\bf X}$ with $({\bf
  Z\;X})$. This yields the following expression for the lower bound:
\begin{equation*}
|\Sigma_0|^{1/2} \Pr(y \,|\, {\bf X}, {\bf Z}) \geq |\hat{\Sigma}|^{1/2}
\textstyle\int e^{g^{\ast}(\beta; \eta)} \, \Pr(\beta) \, d\beta,
\end{equation*}
in which we define
\begin{equation*}
g^{\ast}(\beta;\eta) = \sum_{i=1}^n \log\sigma(\eta_i) 
+ \textstyle \frac{\eta_i}{2}(d_i\eta_i - 1)
+ \hat{y}^T{\bf X}\beta - \frac{1}{2}\beta^T{\bf X}^T\!\hat{D}{\bf X}\beta
+ \frac{1}{2}\hat{u}\hat{\Sigma}^{-1}\hat{u},
\end{equation*}
and we introduce the following notation:
\begin{align*}
\hat{\Sigma} &= (\Sigma_0^{-1} + {\bf Z}^T\!D{\bf Z})^{-1} \\
\hat{u}      &= \textstyle \hat{\Sigma}{\bf Z}^T(y - \frac{1}{2}) \\
\hat{D}      &= D - D{\bf Z}\hat{\Sigma}{\bf Z}^T\!D \\
\hat{y}      &= \textstyle (I - D{\bf Z}\hat{\Sigma}{\bf Z}^T)(y-\frac{1}{2}).
\end{align*}

{\bf Introducing the fully-factorized variational approximation.}
Similar to the linear regression case, the fully-factorized
approximating distribution yields an analytic expression for the lower
bound to the marginal log-likelihood:
\begin{align}
  \textstyle \frac{1}{2}\log&|\Sigma_0| +
  \log \Pr(y \,|\, {\bf X}, {\bf Z}, \theta) \nonumber \\
& \geq {\textstyle \frac{1}{2}\log|\hat{\Sigma}|
+ \frac{1}{2}\hat{u}^T\hat{\Sigma}^{-1}\hat{u}}
+ \sum_{i=1}^n \log\sigma(\eta_i) 
+ \textstyle \frac{\eta_i}{2}(d_i\eta_i - 1)
+ \hat{y}^T{\bf X}r - \frac{1}{2} r^T{\bf X}^T\!\hat{D}{\bf X}r
\nonumber \\
& \quad
- \frac{1}{2} \sum_{i=1}^p ({\bf X}^T\!\hat{D}{\bf X})_{ii}
  \mathrm{Var}[\beta_i]
+ \sum_{i=1}^p \frac{\alpha_i}{2} \bigg[
1 + \log\bigg(\frac{s_i^2}{\sigma_a^2}\bigg)
  - \frac{s_i^2 + \mu_i^2}{\sigma_a^2} \bigg]
\nonumber \\
& \quad
- \sum_{i=1}^p \alpha_i \log\Big(\frac{\alpha_i}{\pi_i}\Big)
- \sum_{i=1}^p (1-\alpha_i) \log\Big(\frac{1-\alpha_i}{1-\pi_i}\Big).
\label{eq:lower-bound-logistic-with-Z}
\end{align}
As before, $\mathrm{Var}[\beta_i]$ is the variance of $\beta_i$ with
respect to the approximating distribution, and $r$ is a column vector
with entries $r_i = \alpha_i \mu_i$.

Finding the best fully-factorized distribution amounts to adjusting
the free parameters $\theta$ to make the lower bound as tight as
possible. The co-ordinate ascent updates for the free parameters are
derived by taking partial derivatives of the lower bound, setting
these partial derivatives to zero, and solving for $\theta$. This
yields the following updates:
\begin{align}
\mu_i &= \textstyle s_i^2 \big(({\bf X}^T\hat{y})_i - \sum_{j \neq i}
({\bf X}^T\!\hat{D}{\bf X})_{ij} \alpha_j \mu_j \big)
\label{eq:update-mu-logistic} \\
s_i^2 &= \big(({\bf X}^T\!\hat{D}{\bf X})_{ii} + 1/\sigma_a^2\big)^{-1}
\label{eq:update-s-logistic} \\
\frac{\alpha_i}{1-\alpha_i} &= \frac{\pi_i}{1-\pi_i} \times
\frac{s_i}{\sigma_a} \times e^{\mu_i^2/(2s_i^2)}.
\label{eq:update-alpha-logistic}
\end{align}
The co-ordinate ascent algorithm consists of repeatedly applying these
updates until a stationary point is reached.

As in the linear regression case, we derive an approximate EM
algorithm to fit the prior variance parameter $\sigma_a^2$. (Recall,
$\sigma^2$ is not needed for logistic regression.) The M-step update
for $\sigma_a^2$ is identical to Equation~\ref{eq:update-sa} after
setting $\sigma^2 = 1$.

{\bf Adjusting the linear approximation to the logistic regression
  likelihood.} In the fourth and final part, we explain how we adjust
the parameters $\eta = (\eta_1, \ldots, \eta_n)$ so that the lower
bound on the marginal likelihood is as tight as possible. The
algorithm is derived interpreting the situation within an EM
framework: in the E-step, we compute expectations (the mean and
covariance of $\beta$); and in the M-step, we maximize the expected
value of the lower bound to the log-likelihood.

We begin by considering the simpler case when we have a single set of
variables $X$. Afterward, we substitute to introduce the additional
variables $Z$. Taking partial derivatives of $E[g(\beta; \eta)]$ with
respect to the variational parameters, we obtain
\begin{equation*}
\frac{\partial E[f(\beta;\theta)]}{\partial \eta_i} =
\frac{d_i'}{2}(\eta_i^2 - (x_i^T\mu)^2 - x_i^T\Sigma x_i),
\end{equation*}
where $x_i$ is the $i$th row of ${\bf X}$, and $\mu$ and $\Sigma$ here
are posterior mean and covariance of $\beta$ computed in the
E-step. The typical approach is to set the partial derivatives to zero
and solve for $\eta$. At first glance, this does not appear to be
possible. But a couple of observations will yield a closed-form
solution: first, the slope $d$ is symmetric in $\eta$, so we only need
to worry about the positive quadrant; second, for $\eta > 0$, $d$ is
strictly monotonic as a function of $\eta$, so $d'$ is never zero.
Therefore, we can solve for the fixed point:
\begin{align}
\eta_i^2 = (x_i^T\mu)^2 + x_i^T\Sigma x_i.
\label{eq:update-eta}
\end{align}

To derive the M-step update for the fully-factorized variational
approximation, after analytically integrating out the coefficients
$u$, we need to replace $\mu$ and $\Sigma$ by the correct mean and
covariance of $\binom{u}{\beta}$ under the variational approximation.
The means and variances of the coefficients $\beta$ are easily
obtained from the variational approximation. The remaining means and
covariances in Equation~\ref{eq:update-eta} are
\begin{align*}
E[u] &= \textstyle \hat{\Sigma}{\bf Z}^T(y - \frac{1}{2} - D{\bf X}r) \\
\mathrm{Cov}[u] &= \hat{\Sigma} + \hat{\Sigma} {\bf Z}^T\!D{\bf X}
                   \mathrm{Cov}[\beta]{\bf X}^T\!D{\bf Z}\hat{\Sigma} \\
\mathrm{Cov}[u,\beta] &= -\hat{\Sigma}{\bf Z}^T\!D{\bf X}\mathrm{Cov}[\beta].
\end{align*}
Therefore, the final M-step update for $\eta$ is
\begin{align}
\eta_i^2 = 
\big(z_i^TE[u] + {\textstyle \sum_{j=1}^p x_{ij} E[\beta_j]}\big)^2
+ z_i^T\mathrm{Cov}[u] z_i + \sum_{j=1}^p x_{ij}^2 \mathrm{Var}[\beta_j]
+ 2 z_i^T \mathrm{Cov}[u, \beta] x_i,
\label{eq:update-eta-fully-factorized}
\end{align}
in which $z_i$ is the $i$th row of ${\bf Z}$.

\end{document}